# STRUCTURAL DYNAMICS AND EVOLUTION OF CAPSULE ENDOSCOPY (PILL CAMERA) TECHNOLOGY IN GASTROENTEROLOGISTS ASSERTION

Maxwell Scale Uwadia Osagie[1], Osatohanmwen Enagbonma[2] and Amanda Iriagbonse Inyang[2]

[1,2]Department of Physical Sciences, Faculty of Science,
Benson Idahosa University,
P.M.B 1100, GRA, Benin City, Edo State, Nigeria.

*ABSTRACT*

*This research paper examined and re-evaluates the technological innovation, theory, structural dynamics and evolution of Pill Camera(Capsule Endoscopy) technology in redirecting the response manner of small bowel (intestine) examination in human. The Pill Camera (Endoscopy Capsule) is made up of sealed biocompatible material to withstand acid, enzymes and other antibody chemicals in the stomach is a technology that helps the medical practitioners especially the general physicians and the gastroenterologists to examine and re-examine the intestine for possible bleeding or infection. Before the advent of the Pill camera (Endoscopy Capsule) the colonoscopy was the local method used but research showed that some parts (bowel) of the intestine can't be reach by mere traditional method hence the need for Pill Camera. Countless number of deaths from stomach disease such as polyps, inflammatory bowel (Crohn"s diseases), Cancers, Ulcer, anaemia and tumours of small intestines which ordinary would have been detected by sophisticated technology like Pill Camera has become norm in the developing nations. Nevertheless, not only will this paper examine and re-evaluate the Pill Camera Innovation, theory, Structural dynamics and evolution it unravelled and aimed to create awareness for both medical practitioners and the public.*

*KEYWORDS*

*Endoscopy Capsule, Inflammatory, Medical, Disease and biocompatible*

## 1. INTRODUCTION

Endoscopy is a medical terminology and refers to instrument used in stomach examination. it is basically for medical treatments and it help known more about the interior of a hollow organ within the stomach. More often than not, it is made up of a video camera with fibre optic tube inserted into the organ and it is popularly known to be gastrointestinal endoscopy (GI). Over the years, this method has not only caused pain and discomfort but has failed to solve the problem associated with small intestine disease because the tube finds it difficult to move around some interior parts of the small intestine for possible bleeding.

It is true that ultrasound innovation came 20 years after the endoscopy but the innovation had it trace to the endoscopy in 1960s [1]. The ultrasound was a solution to the internal GI structures. Advancement in this technology brings about the Endoscopy Ultrasound (EUS) which thus help to determine the level of spread of tumours in the body. The instrument can be use to take tissues sample using Fine Needle Aspiration biopsy (FNA). Research had it that Endoscopy Retrograde Cholangiopancreatography(ERCP) has been in existence for over 28 years using X-rays and endoscopy to know the right state of the body pancreas, gallbladder, ducts and the liver. The X-ray enhances the picture clarity of the associated tiny ducts. According to [1] the scope of ERCP's has expanded, medical centres, like hospital's Therapeutic Endoscopy and GI centres uses it to place tents within bile ducts which help to remove difficult bile duct tones by obtaining biopsy samples.

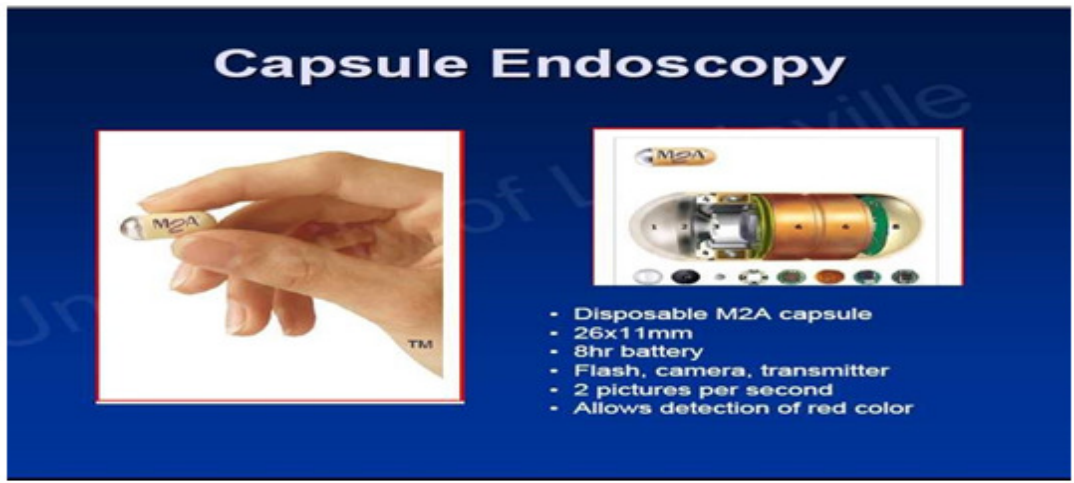

Fig. 1. Pill Camera [4]

Another instrument that has help before the discovering of Pill Camera is the Manometer use by gastroenterologists ensuring proper record of muscle pressure along the oesophagus or anorectic organ. The instrument enables motility stability and this has to do with difficulty in excreting and swallowing. These difficulties are usually associated with acholasia and fecal. Both have sphincter muscle relaxation and stomach disorder due to constipation such as rectal outlet obstruction [2]. One of the problems with the progressive improvement of the different endoscopy instruments in ascertaining diseases and bleeding associated with small intestine is the time duration. It takes several hours and this translates to discomfort on the part of the patient. The most troublesome with the normal endoscopy is the inability to show real image of the affected area of the small intestine. So, not been able to examine critical region within the intestine brought about the tubeless technology (device) embedded with camera that track all critical region within the small intestine.

The device has demonstrated in figure 1. Has a size of a pill [3]. Research had it that the first publication on the findings on Capsule Endoscopy motility in the clinical setting were carried out by [14] and establish that Capsule Endoscopy is helpful for diagnosing patients with irritable bowel syndrome. Indeed the capsule cannot obtain biopsies, fluid such as aspirate and brush lesion for cytology in the small bowel, so the need for real time viewer as well as radio-controlled triggering and remote controlled capsule manipulation for precision is highly recommended. This made it easy for visualization making the capsule miniature laboratory where all gadgets are seen as entity [15; 24].

Research has shown that capsule endoscopy has gone beyond small bowel (SB),the innovative scope currently witnessed brings to the understanding of end user its ability to exploit different and new area in the esophageal capsule (PillCam ESO™) and colonic capsule (PillCam Colon™). Furthermore, it will be loadable to see more research exploitation in Capsule Endoscopy (CE). According to [13] the colonic capsule could be seen as antidote for colorectal cancer monitor and could be traced to the fact that is possesses the non-insidious character. It has been revealed from empirical study that the colonic capsule had two cameras on both ends on first generation and with the capability of absorbing eight frames in two seconds. It is a about 5.1 mm longer when compare to small bowel capsule. The colon must be clean and this helps make no room for remnants of stool so as to reduce associated drawback.

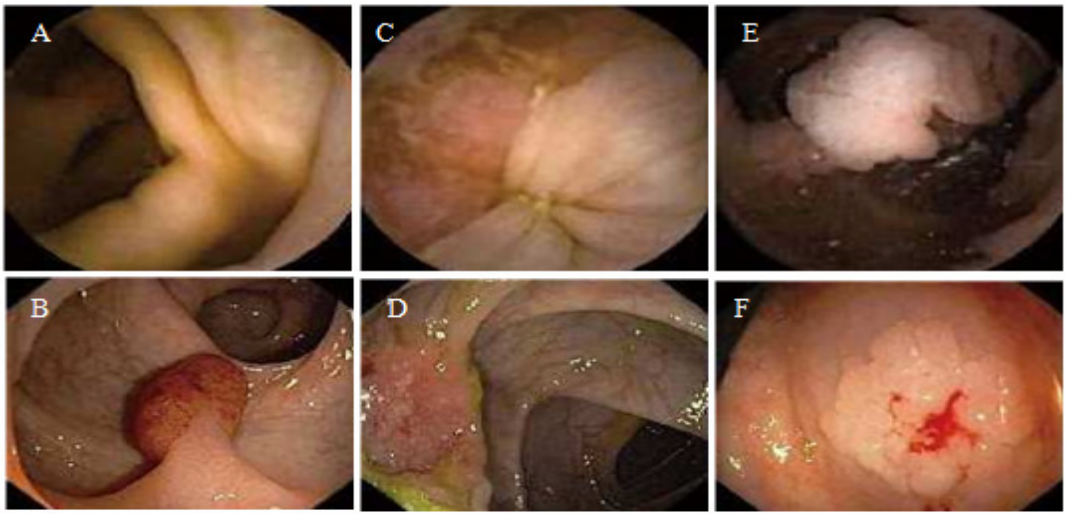

Fig. 2. Images captured by the Pillcam™ Colon and conventional colonoscopy.

A cross session of Figure 2 shows (A-B) as Pedunculated polyp in the sigmoid colon,(C-D) as Ulcerated tumour in the transverse colon, (E-F) as Flat adenoma in the ascending colon. Rather than have these examinations separately it would be idea to have them captured by single capsule thereby leaving the physician to do the rest analysis

A patient should avoid magnetic fields like magnetic resonance imaging (MRI), and metal detectors during capsule examination in the stomach, and this occurs within 24-48 hours. For purpose of clarity, Small bowel preparation is still been seen as critical issue and this make the adaptation of fasting or clear liquids for 10 to 12 hours some could even be as far as 24 hours before the study, however, research shown that bowel preparation could be carried out with 2 to 4 litres of polyethylene glycol based electrolyte solution or oral sodium phosphate preparation with sole responsibility of giving a clear visualization of the small intestine [24].The esophageal capsule (PillCamTM ESO) was approved by the Food and Drug Administration(FDA) in November 2004, after cross-examination which prove the safety of the capsule. Figure 3 and 4 shows pictures of a patient with erosive esophagitis

International Journal in Foundations of Computer Science & Technology (IJFCST) Vol.8, No.1/2, March 2018

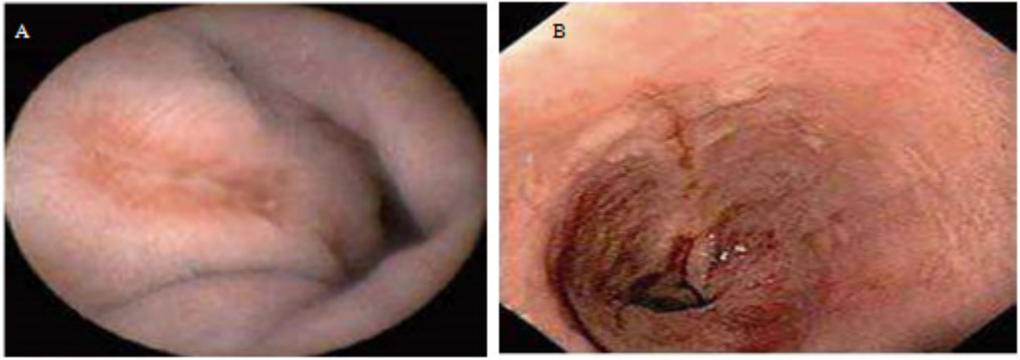

Fig. 3.PillCam™ ESO image of erosive esophagitis; B: endoscopy image of distal esophagus in the same patient.

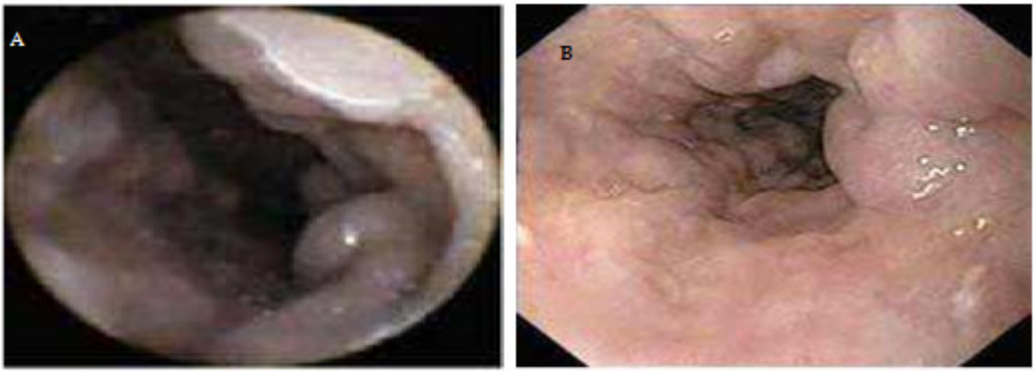

Fig.4.PillCam ESO™ image showing esophagealvarices; B: Upper endoscopy image of distal esophagus in the same patient.

## 2. RELATED LITERATURES

Technology improvement in medical has so far taken new dimension

The technological concept for small bowel capsule as seen in Figure 1 is the brain child of two distinguished inventors,Dr. Paul Swain and Dr. Gavriellddan.Swain, a British gastroenterologist, in 1996 he performed the first live text transmissions using a pig's stomach for the live broadcast. A year later there was synergy Gavriellddda who by profession and training was mechanical engineer Defense Ministry of Israel [5; 6; 7]. Following the text result evaluation the work was made know to the public through publication in 2000 [7]. Though this never came to reality same year published but the first use of capsule edoscopy (CE) was seen in the clinical demonstration in the year 2001 and was widely published in the same year [7] and according to [8] more than 1,000,000 capsules have been swallowed worldwide and nearly 1000 peer reviewed publications have appeared in different literatures.

According to research carried by [9] it was revealed that Pill Camera (Capsule endoscopy) is classified as a new down in the diagnostic tool in medical examination of patient which allow a direct penetration as visual examination of the small intestine, which in case surpassthe area of the previous endoscopy/colonoscopy. The M2A Capsule Endoscopy known to be the Pill has the shape of a multivitamin, this allow is to be able to go through the easophagoeus to the small intestine by the help of water and as said previously it is made up sealed biocompatible material that made it resistant to stomach acid and other enzymes in the body. The biocompatible material goes as far preventing the M2A Capsule from rupture hence, making it not harmful to the body organs. Doctors sees this method as more convenient in reaching the interior parts (bowel) of the body that couldn't be reach by previous or traditional upper endoscopy or by colonoscopy. It is a well known fact that bleeding of the small intestine is the main reason the Pill Camera was invented but it uses could be seen as going far to check for polyps

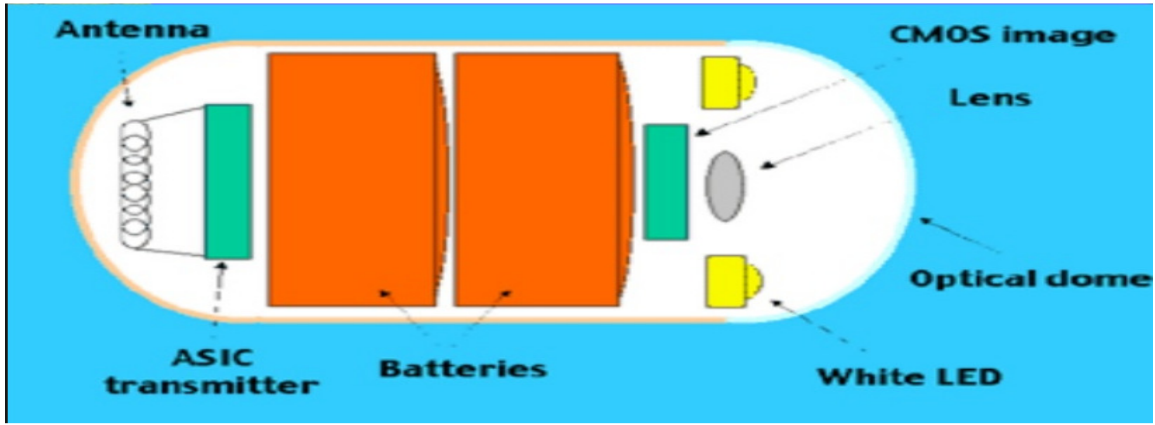

Fig. 5. Pill Camera Architecture [11]

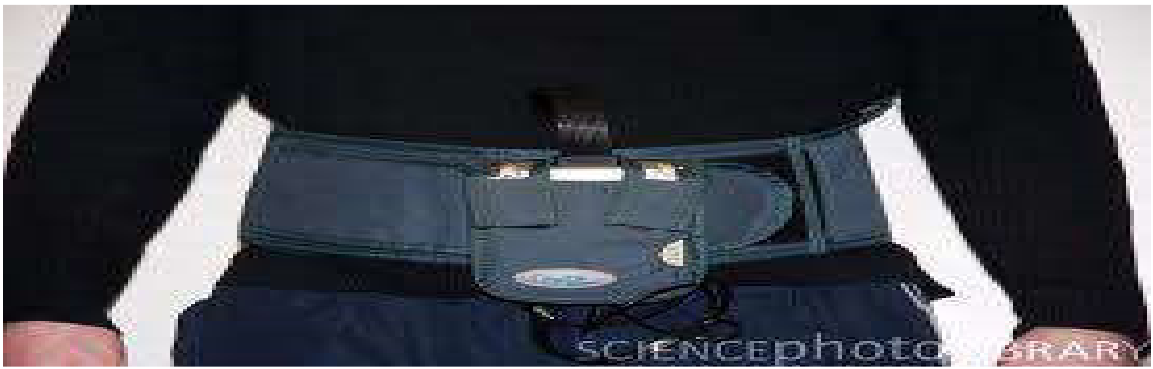

Fig. 6. Pill Camera Data recorder [11]

Mohammed in his research explained that the Pill Camera has seven (7) optical fibres as shown in Figure 5. He further made known that six (6) are used for collecting light and one (1) for illumination. At the point of moving from one region of the body to another once swallowed it create room for electric current flows and this cause the encased fibres to bounce back and forth to make the electronic eye scan the gastrointestinal endoscopytract. The Pill has an embedded red, green and blue illumination laser light that help in visualization of the critical bleeding region. The Pill Camera is designed to capture on two-dimensional which help in the patient diagnosis. The images are directly retrieved from the recording device displayed by Figure 3 and it is worn on patient's waist as a belt [10].

International Journal in Foundations of Computer Science & Technology (IJFCST) Vol.8, No.1/2, March 2018

The increase advancement in the development of modern technology is the positivism on the usage and application of Pill Camera to the medical field. Endoscope since inception has moved from low pace to an advanced state in the diagnoses and examination of the gastrointestinal tract of a patent condition. The initial process of endoscopy is through the insertion of 8mm tube through the mouth, with camera at the tale end, images display on monitor. This method is simply to make the medical practitioners (medics) carryout the gullet passage of the tube down to the stomach [12].Notwithstanding, this process has been classified as cumbersome and pain tasking on both parties. It is no doubt that the tubeless method of the endoscopy has led to evolution of new area awaiting exploitation in research community, this gives clear indication that patient suffering stomach ailment or possible gastrointestinal tract symptom can simply swallow Pill Camera/Endoscopy Capsule that takes snapshots of all internal parts of the body for easy evaluation and examination on the part of the heath practitioners.

## 3. PILL CAMERA AND ITS ARCHITECTURE

As demonstrated in Figure 2., the Capsule is 11 by 26 mm (11x26 mm) with outer isoplast envelope[23]. This made it biocompatible and unreceptive stomach enzymes or gastric liquid. The capsule designed has some elements that made it unique and irrespective the miniature it has light emitting diode (LED) with an embedded lens, two (2) batteries known to be silver-oxide, a microchip camera, transmitter with and antenna radiation and automatic sensor switch. The beauty of the camera is on it low power consumption because it is built with complementary metal oxide semiconductor technology. The batteries power the CMOS detector, led and transmitter for easy visibility of the small intestine tissues. The lights as distinguished previously have major role in determining the extent the diagnoses can go because medical practitioners like pathologist does the disease detection by colour. The CMOS detector is embedded with signal to noise ratio (SNR), light emitting diode (LED) and the Application Specific Integrated Circuit (ASIC), these three elements enable the visibility of the tiny image, theses elements are inseparable and remain an entity. It is observed that there is an improvement in the development of ASIC and this has metamorphose into efficient video transmission of bandwidth as well as power with minimum amount into the capsule hence synchronization of LED, CMOS and ASIC consumption. The captured images after excretion are taken to computer for onward analysis. These computers are vehemently with the suitable software tool that help in the diagnoses and give room for medical expert such as the physician to have clear understanding of the effect ratio of the affected tissue or tissues in the body

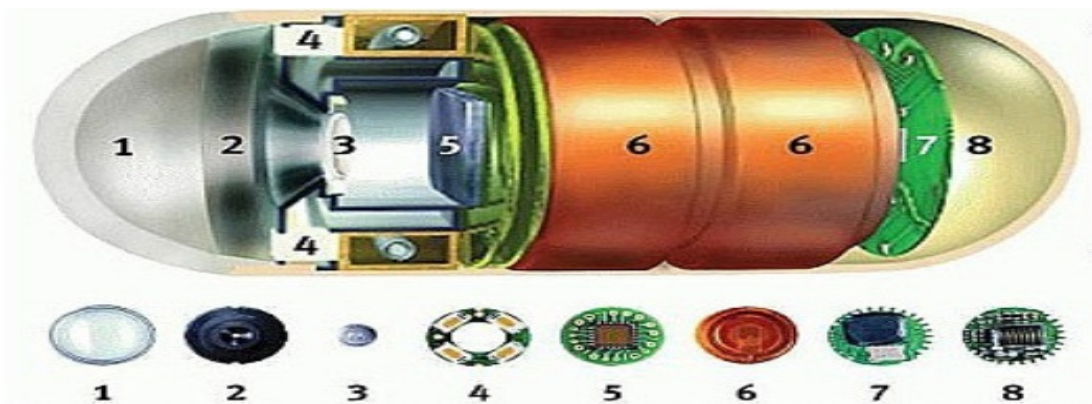

Fig. 7. Internal mechanism of a Pill Camera (capsule) [11]

Table I. Dimensions of the Internal Mechanism of Pill Camera

| Serial Number | Features OF Pill Camera | Functions |
| --- | --- | --- |
| 1 | Optical Dome | It is a non conductive element of the capsule and represents the front of the capsule shape. It functions involved liquid filtration and other enzymes balancing |
| 2 | Lens Holder | As the name implies it holds the lens firmly to the capsule to prevent detachment |
| 3 | Lens | The lens is the eye of the capsule and can't be underestimated. It formed one of the major parts of the Capsule and it is place under the side of the Optical Dome |
| 4 | LED | There are several components that made up of the LED and the end product of it is to illuminate light around the passage region of the body for easy identification of affected tissues. It prevent reflection through the light receiving window |
| 5 | CMOS Sensor | This is the beauty of the Capsule because it detect minutes object as less than 0.2 mm and work on a precision of 140 degree |
| 6 | Battery | The design nature of the battery made it harmless to the body and as explained previous it is made up of silver oxide. It is two to make it last for the period of examination and has a button shape |
| 7 | ASIC Transmitter | The transmitter has two electrodes isolated electronically and it is an integrated circuit which help the facilitation of images captured |
| 8 | Antenna | The communication existing between the belt receiver and the capsule is done by the help of antenna which a coated polyethylene |

## 4. COMMUNICATION COMPONENTS

The capsule as explained previously does not have the sole responsibility of performing the diagnostic analyses, other components work simultaneously with the capsule to bring about holistic diagnosis in patient suspected to be having the bleeding of the small intestine. Patient undergoing the endoscopy examinations are meant to put on the varieties of components associated with the capsule endoscopy before readings and evaluations are made. So, to make the images transmitted to solve the problem, proper record must be taken. A is meant to wear an antenna with wire connected to the recording unit. The antenna array deigned to be worn under normal clothing is put right to the chest of the patient under electrocardiography and multiple numbers of images transmitted by the capsule, received is recorded. It has no stationary part with the patient because study or research has not shown the endemic problem associated with body movement during cross examination and evaluation of patient under Capsule endoscopy text.

a. Pill Cam Capsule: SB/ESO: as approved by Food and Drug Administration (FDA) are placed in tabular form

Table II. Pill Cam Capsule Recommended

| SB | ESO |
|---|---|
| For small bowel. | For esophagus. |
| Standard lighting control. | Automatic lighting control. |
| One side imaging. | Two sided imaging. |
| Two images per second. | 14 images per second. |
| 50,000 images in 8 hours. | 2,600 images in 20 minutes. |

b. Sensor Array Belt.

The sensor array belt is more or less like the Electrocardiogram (ECG) used in obtaining the movement pattern of the heart. This is seen as a wave length or movement indicating the pattern of the functionality of the heart. The wires are directly join to light weight data recorder. The sensor array help monitor the rightful position of the capsule within the body. A belt is worn around the waist with recording device and a battery pack. The sensor encompassa pad sensors, battery, cable, and bag receiver.

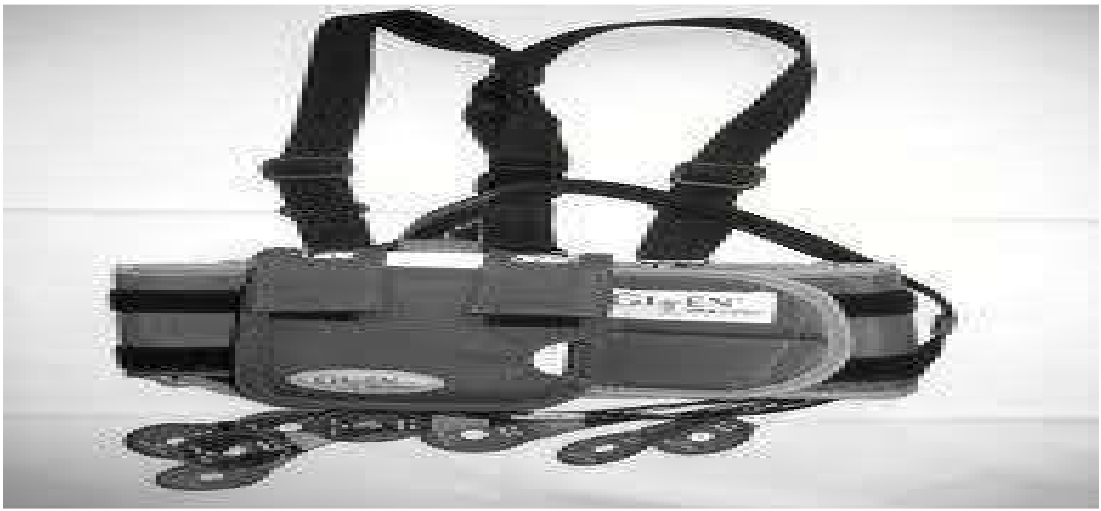

Fig.8. Sensor array belt [11]

c. Data Recorder:

The recorder is attached to the sensor belt and is portable enough to fit into the recorder pouch with little above 489 gm of weight. The signal transmitted by the camera through sensory array are captured this device. It has ability to store images of 5500 to 6000 JPEG format on a drive capacity of about 10 GB. The images download speed is directly proportional to the network available on processing time and this shown by the belt worn around the waist captured by Figure 6. On usage, the sensory array, recorder belt and the battery must be orderly and neatly done to avoid damage or some environmental hazard capable of affecting the device

d. Real Time Viewer:

The time viewer is a Liquid-crystal Display (LCD) with embedded rapid reader software as seen in Figure 9. It is a real time viewing machine that make know the view the current position of the capsule within the body

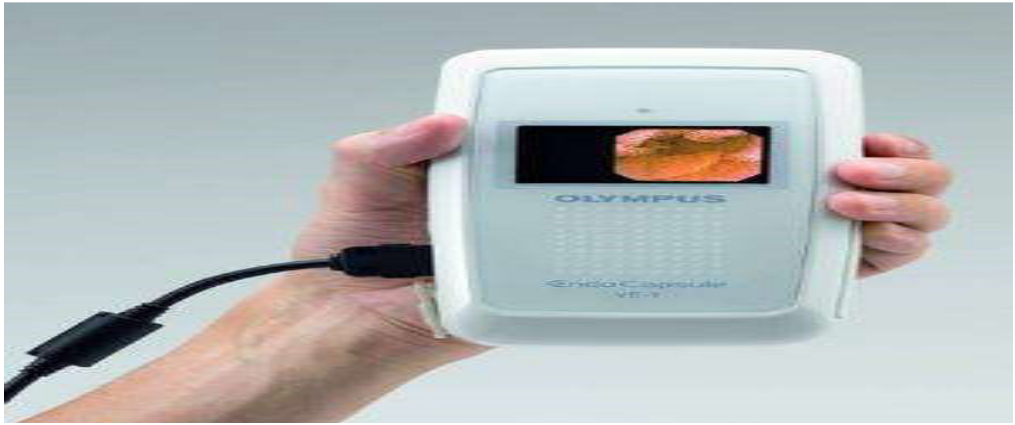

Fig. 9. Real time viewer [11]

e. Work Station and Rapid Software.

The work station is the centre for rapid processing and a base for feedback of images receive from through different elements responsible. The downloaded images are further process by Rapid Application Software (RAS). The medical practitioner do the analysis by watching the two dimensional images. The software application has since be improved on through features such degree of localization capsule around the abdomen inline with the video images. it also has the ability to automatically point out the capsule images that cohabits with the already existing blood or red sport

## 5. ENDOSCOPY (PILL CAMERA) EVALUATIONS AND DISCUSSIONS

It is difficult to have a system or innovation void of drawbacks, it is same with capsule endoscopes and this solvable solution is indeed the true reflection of what happen in the next feature in microelectronics capable of creating image sensors with a smaller pixel size and higher resolution. Capsule Endoscopy as practiced uses image data compression which causes blot at to objects thereby leading to lower image quality. This is a serious limitation to the capsule endoscopy innovation. Another setback is the diminution of the two silver oxide batteries used in the current capsule endoscopy which could cause complete imaging of the small intestine when kept for longer time in the stomach [16; 17]. Though, this setback could form less penetration in critical research into capsule endoscopy other researchers such [18; 19] found out solutions to the prevalent problem of long time examination. Two research on capsule endoscopy are currently being supervised by the European Union (EU), the VECTOR (Versatile Endoscopy Capsule for gastrointestinal Tumor Recognition and therapy) and NEMO (Nano based Capsule-Endoscopy with Molecular Imaging and Optical biopsy)[20; 21; 22]. The successful delivering on this will mark the new down to the ever present capsule endoscopy.

According to [25]the procedure set aside by the American Society of Gastrointestinal Endoscopy (ASGE) on platelets count cirrhosis and cholestatic liver disease has brought about the esophago-gastro-duodenoscopy(EGD). In the cross session, grades 1 to 3 were evaluated: C0 = no varices, C1 =small and nontortuousvarices<25% of the circumference of the frame, and C2 =large varices>25% of the frame circumference. Similarly, [26]show that multicenter international study with PillCamesophageal prior to esophago-gastro-duodenoscopy was performed in 97 cirrhotic patients within 48 hours by endoscopists sightless the results of capsule endoscopy, while the PillCamesophageal study was read by a sightless second investigator making the sensitivity a little above 86.5% and 86.6%, respectively and this could be seen as perfect result.

Again, the statistic published by [27] show a study of 328 patients, and the capsule endoscopy sensitivity in actually detecting polyps ≤ 6 mm in size were 64% (95% confidence interval 59–72) and 84% (95% CI 81–87), respectively, and for detecting advanced adenoma sensitivity and specificity were 73% (95% CI 61–83) and 79% (95% CI 77–81) respectively of 19 cancers detected by colonoscopy, 14 were detected by capsule endoscopy (sensitivity 74%, 95% CI 52–88). For all lesions, the sensitivity of capsule endoscopy was higher in patients with good or excellent colon cleanliness compared to those with fair or poor colon cleanliness. This again could be classified as a desire result worthy enough to change the capsule endoscopy research to positive state. As demonstrated by [28], 79% sensitivity and 94% specificity of capsule endoscopy for Barrett's esophagus in 77 patients is a way forward to the actualization of the novel full scale.

It is true that the statistical result of the various test conducted speaks volume but it would have been better to see clear collaboration with other researchers to ascertain it validity and current literatures on capsule endoscopy.

## 6. CONCLUSION

Wireless capsule endoscopy is indeed a breakthrough in small bowel investigation. Following the drawbacks associated with other available techniques to image this region capsule endoscopy has bridged the gap. This innovation in the nest couple of years will increase it scope in wide range of patients with variety of illnesses. It could be said that the innovation suited for patients with gastrointestinal bleeding of unclear etiology. The innovation ideal detection of small lesions caused by bleeding such as tumours and ulcers makes it the right instrument for quick and easy discovering.  Although, there are wide variety of indications for capsule endoscopy being investigated currently but the capsule pill camera innovation offers a comparative advantage over them. The capsule endoscopy is painless and effection free, it offers advantage such as miniature size, accurate, precise (view of 140 degree), high quality images, harmless material, simple procedure, high sensitivity and specificity, avoidance of risk in sedation and efficient than X-ray, CT-scan, and normal endoscopy. Capsule endoscopy application is across medical industry, robots, sophageal diseases, gastrointestinal reflex diseases, barreff's esophagus, Crohn's disease, small bowel tumours, small bowel injury, celiac disease, ulcerative colitis etc. The future is bright for capsule endoscopy because it will become effective in diagnostic gastrointestinal. This innovation makes patients with cancer or varicesless worried because it has easy and painless procedure when compared to conventional colonoscopy and gastroscopy. The VECTOR and NEMO research being sponsored by the EU will be another interesting area in the near future.

International Journal in Foundations of Computer Science & Technology (IJFCST) Vol.8, No.1/2, March 2018[16] Triester, S. L., Leighton, J. A., Leontiadis, G. I., Fleischer, D. E., Hara, A. K., Heigh, R. I., ... & Sharma, V. K. (2005). A meta-analysis of the yield of capsule endoscopy compared to other diagnostic modalities in patients with obscure gastrointestinal bleeding. The American journal of gastroenterology, 100(11), 2407-2418.

[17] Moglia, A., Menciassi, A., & Dario, P. (2008). Recent patents on wireless capsule endoscopy. Recent Patents on Biomedical Engineering, 1(1), 24-33.

[18] de Franchis, R., Rondonotti, E., Abbiati, C., Beccari, G., & Signorelli, C. (2004).Small bowel malignancy. Gastrointestinal Endoscopy Clinics, 14(1), 139-148.

[19] Liao, Z., Li, Z. S., &Xu, C. (2009). Reduction of capture rate in the stomach increases the complete examination rate of capsule endoscopy: a prospective randomized controlled trial. Gastrointestinal endoscopy, 69(3), 418-425.

[20] Gao, M., Hu, C., Chen, Z., Zhang, H., & Liu, S. (2010). Design and fabrication of a magnetic propulsion system for self-propelled capsule endoscope.IEEE Transactions on Biomedical Engineering, 57(12), 2891-2902

[21] Valdastri, P., Quaglia, C., Buselli, E., Arezzo, A., Di Lorenzo, N., Morino, M., ...& Dario, P. (2010). A magnetic internal mechanism for precise orientation of the camera in wireless endoluminal applications. Endoscopy, 42(06), 481-486.

[22] Swain P, Toor A, Volke F, Keller J, Gerber J, Rabinovitz E & Rothstein RI.(2010). Remote magnetic manipulation of a wireless capsule endoscope in the esophagus and stomach of humans (with videos).GastrointestEndosc 71, 1290-1293

[23] Davis, B. R., Harris, H., & Vitale, G. C. (2005). The evolution of endoscopy: wireless capsule cameras for the diagnosis of occult gastrointestinal bleeding and inflammatory bowel disease. Surgical innovation, 12(2), 129-133.

[24] Dai, N., Gubler, C., Hengstler, P., Meyenberger, C., &Bauerfeind, P. (2005). Improved capsule endoscopy after bowel preparation. Gastrointestinal endoscopy, 61(1), 28-31.

[25] Qureshi, W., Adler, D. G., Davila, R., Egan, J., Hirota, W., Leighton, J., ...& Baron, T. H. (2005). ASGE Guideline: the role of endoscopy in the management of varicealhemorrhage, updated July 2005. Gastrointestinal endoscopy, 62(5), 651-655.

[26] Eisen G. (2006). Esophageal capsule.Presented at the ICCE meeting Boca Raton, Abstract 20154. FA, USA.

[27] Van Gossum, A., Munoz-Navas, M., Fernandez-Urien, I., Carretero, C., Gay, G., Delvaux, M., ...&Costamagna, G. (2009). Capsule endoscopy versus colonoscopy for the detection of polyps and cancer.New England Journal of Medicine, 361(3), 264-270.

[28] Galmiche, J. P., Sacher-Huvelin, S., Coron, E., Cholet, F., Soussan, E. B., Sébille, V., ...& Le Rhun, M. (2008). Screening for esophagitis and Barrett's esophagus with wireless esophageal capsule endoscopy: a multicenter prospective trial in patients with reflux symptoms. The American journal of gastroenterology, 103(3), 538-545.
12